# Hybrid Cloud Security: Balancing Performance, Cost, and Compliance in Multi-Cloud Deployments

**Anjani kumar Polinati**

*M.C.A., Andhra University, India*

Senior Software Engineer, Independent Researcher

Wesley Chapel, FL, USA

Email: anjanikumarpolinati@gmail.com

ORCID: https://orcid.org/0009-0005-0165-0675

**Abstract**

The pervasive use of hybrid cloud computing models has changed enterprise as well as Information Technology services infrastructure by giving businesses simple and cost-effective options of combining on-premise IT equipment with public cloud services. hybrid cloud solutions deploy multifaceted models of security, performance optimization, and cost efficiency, conventionally fragmented in the cloud computing milieu. This paper examines how organizations manage these parameters in hybrid cloud ecosystems while providing solutions to the challenges they face in operationalizing hybrid cloud adoptions. The study captures the challenges of achieving a balance in resource distribution between on-premise and cloud resources (herein referred to as the "resource allocation challenge"), the complexity of pricing models from cloud providers like AWS, Microsoft Azure, Google Cloud (herein called the 'pricing complexity problem'), and the urgency for strong security infrastructure to safeguard sensitive information (known as 'the information security problem'). This study demonstrates the security and performance management solutions proposed were validated in a detailed case study of adoption of AWS and Azure based hybrid cloud and provides useful guidance. Also, a hybrid cloud security and cost optimization framework based on zero trust architecture, encryption, hybrid cloud policies, and others, is proposed.

The conclusion includes recommendations for research on automation of hybrid cloud service management, integration of multi-clouds, and the ever-present question of data privacy, stressing how those matters affect contemporary enterprises.

Keywords: Data encryption, privacy, cloud and multi-cloud services, zero trust, hybrid and composite clouds, integration, cost and performance, security.

# 1. Introduction

## Models of Computing Clouds

Business and organizational adoption of cloud computing is changing the boundaries of how organizations administer and control the information technology systems and the infrastructure. There is no need for a business to have local IT infrastructure or management for a business to operate since application software, servers, and even storage resources can be made available over the internet. Likewise, the organization implementing cloud computing can become more agile in the scaling of operations while at the same time reducing the costs and complexity associated with traditional IT systems infrastructure. With the passage of time, cloud computing has evolved from classic single clouds models of public and private clouds to more advanced systems termed as hybrid and multi-clouds where an organization uses services from several cloud service providers and multiple deployment models.

The development of hybrid and multi-cloud infrastructures stems from increasing market demands in flexibility, cost-efficiency, and data governance. With a hybrid cloud model, an organization usually combines one or more public cloud infrastructures with a private cloud, allowing data and applications to flow freely across different environments (Buyya et al., 2008). In this way, organizations can take advantage of public clouds for less sensitive processes, while more sensitive and critical processes are done in the private cloud environment. Furthermore, multi-cloud environments simultaneously engage multiple cloud service vendors, which enhances redundancy and mitigates the issues of vendor lock-in, and thus enables greater flexibility and responsiveness. This approach is becoming increasingly popular as organizations unbundle from a single cloud service provider while striving to satisfy performance, compliance, and costs.

**Significance of Hybrid Cloud in Today's Information Technology Sector**

There's little doubt that the transition to hybrid cloud systems is becoming more common these days. Modern organizations are no longer restricted to single-cloud systems; they look for ways to enhance their IT ecosystems by improving performance, cost, and compliance. As Jackson and Jeffery put it," the enhanced hybrid cloud architecture facilitates an organization to scale its IT activities as needed, with the provision of control over the required data and processes." There is a choice for customizing the management approach of operating and workload between private and public cloud environments to better satisfy specific business needs.

Furthermore, in addition to flexibility, hybrid cloud environments provide another feature of cloud computing: scalability. Organizations can automatically increase or decrease their infrastructure IT resources to respond to workload requirements without having to spend a lot of capital on physical equipment. This is especially beneficial to companies subject to fluctuation in computing demand, as it allows upscaling or downscaling in response to market and seasonal transitions or internal business growth (Varghese & Buyya, 2017).

Despite this, the advantages of hybrid clouds are counterbalanced by a specific combination of challenges especially with regard to security and compliance. Safeguarding sensitive information and complying with various regulations' requirements becomes even more difficult with the many cloud environments that exist. There is a risk of data breach and loss of control over data with the integration of many different cloud platforms and, therefore, advanced security event management is required (Gai et al., 2020). Healthcare, finance, and government industries have restricted perimeters which makes compliance another difficult issue to deal with. This model is also flexible, but a great deal of care has to go into planning for compliance with local and international data protection laws and industry standards.

**Table 1: Comparison of Different Cloud Models (Public, Private, Hybrid, Multi-cloud)**

| Cloud Model | Definition | Advantages | Challenges |
|---|---|---|---|
| **Public Cloud** | Cloud services offered over the internet by third-party providers. | Cost-effective, scalability, no infrastructure maintenance. | Limited control over data, security concerns. |
| **Private Cloud** | Cloud infrastructure operated solely for one organization. | Greater control over data, enhanced security. | Higher upfront costs, maintenance responsibility. |
| **Hybrid Cloud** | Combines public and private clouds, enabling data and applications to be shared between them. | Flexibility, scalability, control over sensitive data. | Integration complexity, security and compliance risks. |
| **Multi-cloud** | Use of services from multiple cloud providers. | Avoids vendor lock-in, improved reliability. | Complexity in management, cost management challenges. |

The growing focus on hybrid cloud models Recently, shifts towards the hybrid model of cloud computing have been witnessed, which indicates that emphasis should be placed on adequate security and compliance strategies for multi-cloud deployments. This is one of the issues regarding which this paper will investigate in detail: how to make sure that hybrid and multi-cloud environments are secured while at the same time meeting performance and compliance requirements.

## 2. Literature Review

**Cloud Computing Security Challenges**

The versatility of cloud computing has many benefits, including scaling and reduced costs. Still, businesses that move to the cloud face particular security threats that may compromise their data's confidentiality, integrity, and availability. One significant issue identified by Zhang et al. (2013) and Bernstein et al. (2009) concerns the possibility of data breach incidents, in which sensitive information can be accessed by unauthorized parties. These risks are more pronounced by the shared responsibility approach, where both cloud customers and service providers are responsible for the security of the system, and as a result, do not understand the demarcation of responsibility.

The most prevalent security threats in the cloud are the safeguarding of data in motion and at rest, identity and access control, and the risk posed by cyber attacks. These weaknesses are aggravated in public clouds where intra-organization employees can constitute a risk owing to the absence of physical security oversight from the cloud service provider. Security management becomes even more difficult in multi-cloud environments that make use of several cloud service providers. It is very difficult to maintain homogeneous security policies across different systems because every cloud provider has their own customized security policies. Zhang et al. (2013) argue that the distribution of information in the different cloud systems raises the problem of data privacy and access control. This form of control for cloud computing becomes more complicated with the enforcement of third-party compliance requirements like EU GDPR or HIPAA that have restrictive and prescriptive data governance procedures and policies.

Furthermore, the dispersion of data across different regions poses risks to data sovereignty as companies might accidentally save data in regions with inadequate legal and privacy safeguards. Moreover, the lack of precise access control levels, where only selected users are supposed to obtain sensitive data, creates another challenge in multi-cloud infrastructure environments. Correspondingly, there becomes greater difficulty to ensure the automatic application of consistent security policies due to the many security systems and access control techniques an organization has to contend with.

**Security Issues Related To Hybrid Clouds**

The move to the use of hybrid clouds creates additional security issues that become more complicated. A typical hybrid configuration integrates a private cloud along with a public one. However, from a security perspective, these setups pose a significant threat. These risks have, however, been outlined in multiple for security frameworks that seek to fill the gaps in the security of hybrid clouds. One of the most and the earliest cloud security frameworks was put forward by Fox and his colleagues in 2009. They proposed a layered method towards security of the cloud which included mitigating risks such as encrypting data, enforcing secure access controls, and active surveillance of the cloud. This approach indicates the need for robust security provisions that are capable of efficiently and adequately coping with any form of threat or vulnerability that arises. The framework also underscores the need for clearly defined data policies that restrict the sensitive information to strict measures so that it is not exposed together with less sensitive information.

Maintaining privacy and integrity of data in hybrid cloud environments depends largely on security frameworks being implemented. Even though the hybrid cloud model allows organizations to merge the flexibility of public clouds with the control provided by private clouds, the risk of potential security breaches increases at a phenomenal scale if appropriate measures are not taken. Fox et al. (2009), for example, contend that perimeter security (such as firewalls) cannot be the lone bastion relied upon for security in hybrid clouds, and active directory-based identity management systems alongside secure application programming interfaces (APIs) and strong encryption for data in transit across cloud environments are essential.

A rather paramount security framework undergoing development with regard to hybrid clouds is the concept of Zero Trust architecture. Kindervag (2010) argues that 'Zero Trust' was previously defined by the absence of boundaries. Within that 'boundary-less' framework, systems have to trust that threats try to break in from both within the organization and from the outside, thus, default trusting of an individual on the inside or outside the organization becomes unsafe. This model is very important in a hybrid cloud environment as it offers a complete solution to the problem of verifying and securing any request for access to resources, irrespective of where the request is coming from. Unlike traditional security approaches which operate with the assumption that one internal network is safe, and all that needs to be done is to control and validate external connections. It is achieved through strong identities verification, continuous monitoring and least privileged access of every user and devices in control.

Although Zero Trust models and frameworks like those proposed by Fox et al. (2009) and Kindervag (2010) offer robust security coverage for hybrid clouds, there are some challenges. Effective deployment of these models incurs both high capital expenditure and operational expenditure in technologically advanced resources. Moreover, the small and resource-constrained organizations can be overburdened with the complexity of managing continuous authentication and monitoring. However, as the adoption of hybrid cloud models increases, the need for robust security frameworks to mitigate various threats and vulnerabilities will be increasingly important.

3. **Hybrid Cloud Models and Security Requirements**

**Types of Hybrid Cloud Architectures**

Having noted one concern of modern organizations, the most crucial issue of hybrid cloud computing is the design of the hybrid cloud architecture. Such architectures support the integration of a private cloud and public cloud to allow more flexibility in meeting business needs. As put forth by Armbrust et al. (2009), the public-private hybrid cloud model consists, as the name suggests, of a privately owned unit which is called a private cloud (an organizational unit) and a publicly accessible unit which is called a public cloud (an organizational unit that can be subscribed to by various users). For this type of hybrid architecture, organizations are able to maintain critical data and applications in a private cloud unit while enjoying the flexibility, cost-effectiveness, and scalability of public cloud services. Such hybrid cloud models allow organizations to protect sensitive information while taking advantage of the public cloud's computational capabilities.

The hybrid cloud model involves more than just private and public clouds. It has evolved with technology to include multi-cloud environments that incorporate multiple public cloud services along with private clouds. As cited in Mell and Grance (2011), this allows firms to reduce vendor lock-in, increase service redundancy, and enhance performance by having several cloud providers competing for their business. Unfortunately, these loose environments create new issues for security and control. Organizations must enforce a single set of security policies that span all of the providers because each one may have its own unique features and security protocols. This creates additional barriers to seamless data integration, effective risk management, and compliance. Because there is a growing variety of available cloud environments, organizations need to adopt well-defined and thoughtfully tailored hybrid cloud models that will enable multi-cloud strategies while protecting sensitive information.

**Key Security Issues in Hybrid Clouds**

Hybrid cloud architectures are one of the most attractive models available today, but they come with a host of security problems that must be dealt with. One of the more prominent problems is ensuring data integrity, confidentiality, and secure communications. Sultan (2009) points out that integrity and confidentiality of data is perhaps the most basic issue which needs to be resolved in the hybrid cloud model. Strong encryption methods need to be employed to protect data while being transferred as well as data at rest within the cloud. If this data can be accessed or altered by outside users, it would be damaging to the organization from both a financial and reputational standpoint. Additionally, encrypted communication links must be provided to the on-premise infrastructure to cloud domains, such that the data being communicated is not vulnerable during the transfer process.

When data and workloads are distributed across multiple locations in a hybrid cloud, there is a greater risk of sensitive information being accessed without authorization. According to Mell and Grance (2011), it is essential for organizations to bolster their data protection approaches with end-to-end storage encryption and strict access control policies, meaning the choice of encryption algorithms has to be made with precision to ensure their impermeability to contemporary attacks, for example, brute force or side channel attacks. Moreover, the management of access permissions along with roles is critical in ensuring that unauthorized users do not obtain access to sensitive data, applications, or systems. The control models based on roles and the attributes, respectively, RBAC and ABAC, are widely implemented fostered by the so-called 'minimum use' policy meaning the maximum limitation of user rights to the necessary scope of system resources.

Regulatory compliance poses another challenge for hybrid cloud services. Several industries, such as healthcare, finance, and government, have stringent requirements for data retention, processing and access. Jackson and Jeffery (2015) argue that the design of hybrid cloud models is required to consider the compliance features of these industries which include some frameworks such as General Data Protection Regulation (GDPR), or the Health Insurance Portability and Accountability Act (HIPAA). Such regulations impose restrictions on the geographic storage of sensitive information, its encryption, and the access control mechanisms employed. In a hybrid cloud, the compliance challenge becomes more significant in scope and complexity because more environments, and more cloud service providers, are involved. Complying with these requirements involves paying special attention to the locality of the data and the protective measures taken. Organizations must ensure that both the public and private cloud components of their hybrid architecture comply with these legal and regulatory requirements, which frequently involve the establishment of audit trails, logging data access, and other active measures of compliance.

**Table 2: Security Challenges in Hybrid Cloud Models**

| Cloud Model | Security Challenges | Mitigation Strategies |
|---|---|---|
| **Public Cloud** | Risk of unauthorized access, data breaches, shared infrastructure vulnerabilities, lack of visibility | Strong encryption, identity and access management, regular audits, service level agreements with providers |
| **Private Cloud** | Limited scalability, potential internal threats, complex security configurations | Enhanced access control, encryption, internal monitoring systems, perimeter security |
| **Hybrid Cloud** | Data integrity, privacy concerns, compliance with regulatory standards, inconsistent security policies between public and private clouds | End-to-end encryption, multi-factor authentication, role-based access control, compliance monitoring |
| **Multi-cloud** | Vendor lock-in, inconsistent security across multiple providers, challenges in data integration | Standardized security protocols, integration platforms, unified security management across clouds |

**Hybrid Cloud Security Controls**

To alleviate the risks arising from hybrid cloud adoption, an organization should develop a comprehensive set of hybrid cloud security measures. Such measures are required for the safeguarding of information, software, and voice communications in hybrid clouds. One critical aspect of hybrid cloud security is data encryption, which guarantees that information cannot be read or changed without the encryption key, even when it has been intercepted. According to Buyya et al. (2008), encryption is required to be executed for both data at rest and in transport to be considered confidential and intact. Hybrid clouds employ AES, a commonly used encryption algorithm, to prevent data frombeing compromised when it is transferred between on-premise data centers and cloud facilities.

Besides encryption, securing hybrid cloud deployments also relies on Multi-Factor Authentication (MFA). MFA makes certain that users who access cloud resources authenticate themselves with at least two methods, for instance, a password with either a biometric or a one-time passcode. This adds another layer of security because even if a user has compromised an authentication factor, sensitive information cannot be accessed without additional credentials. Furthermore, identity management systems are crucial for supporting users, devices, and applications' identities in the hybrid cloud infrastructure.

**4. Performance and Cost Considerations in Hybrid Cloud**

**Performance Optimization**

Application of a hybrid architecture requires meticulous performance optimization in terms of cost efficiency like any other cloud infrastructure. The existence of hybrid clouds, which typically feature a combination of on-premise components and wider public and private cloud services, increases the need for perpetual optimization of all aspects of performance. In her study, Reinders (2020) argues that hybrid clouds intelligence comes from workload contouring and application functionality comprehension. Performance is largely determined by the placement and structures of workloads, which means that other areas can be improved if other workload performance goals are reached. For example, HPC (high performance computing) clouds are ideal for many applications, however, some of them are latency sensitive and therefore need to be placed on-prem in order to achieve specified response times.

Another essential component effective strategy is the implementation of load balancing in the hybrid cloud architectures. Load balancing refers to the division of work between cloud and on-premise infrastructure so that resources are fully utilized. It improves cloud application performance and reduces the chances of system bottlenecks. It also helps organizations mitigate the risk of system overload, which leads to poor performance of a single infrastructure component. In addition, systematic monitoring and performance analysis is crucial for recognizing performance degradation issues and for mitigating these problems obligatory before users are affected. According to Varghese and Buyya (2017), such integrations call for strong performance management frameworks that consolidate on-premise and cloud resources while offering comprehensive performance monitoring and system control for effective seamless operations.

**Cost Optimization**

Management of an organization's costs in a hybrid cloud model is necessary for minimum operational and economic resources. Every company must make sure that cost are minimized as Sultan (2009) and Varghese & Buyya (2017) remark that preocupations center around performance. Resources in the cloud must be premeditated so that services provided meet rather than exceed demand. Dynamic resource scaling is a useful cost containment technique which enables an organization to adjust the level of its infrastructure to the level of demand. Organizations thus have the freedom to spend only on services utilized.

The budgetary goals facilitate expenditure minimization within hybrid cloud systems. By anticipating future use, forecasting growth, and analyzing past data, businesses can plan budgets and effectively eliminate unforeseen factors. Resource consumption management tools such as those provided by AWS, Google Cloud, and Microsoft Azure allow businesses to keep track of cloud spending and facilitate monitoring of the services provided. Resource spending, warning and expenditure thresholds, and detailed resource utilization reports can all be monitored by users. Cost management is essential in maintaining budgetary constraints within a hybrid cloud framework.

Familiarizing oneself with the pricing strategies of hybrid cloud service providers is useful, since each cloud providers optimizes costs differently. AWS, Google Cloud, and Microsoft Azure utilize different approaches, for instance, payment-by-value, subscription service, and retainer plans. Take, for example, AWS whose pricing relies heavily on how much compute, storage, and data transfer resources a customer needs for the business to operate. Google Cloud utilizes a similar use-based payment structure, but also offer discounts for sustained and committed usages. Flexible pricing options, including pay-as-you-go and reserved instance pricing, are offered by Microsoft Azure and can be beneficial towards lowering expenditures over time if the right commitments are made. With the options given, enterprises are able to find the most appropriate and cost effective moderately hybrid cloud services for themselves.

**Table 3: Cost Comparison of Hybrid Cloud Solutions (AWS, Google Cloud, Microsoft Azure)**

| Cloud Provider | Pricing Model | Compute Costs | Storage Costs | Data Transfer Costs | Free Tier | Key Advantages | Key Disadvantages |
|---|---|---|---|---|---|---|---|
| Amazon Web Services (AWS) | Pay-as-you-go, Reserved instances, Spot instances | EC2 pricing based on instance type and region | S3 pricing based on usage and region | Data transfer is based on region and amount | Free tier available for 12 months | Extensive service offerings, strong security features | Complex pricing structure, can be expensive for small-scale deployments |
| Google Cloud | Pay-as-you-go, Committed use, | Google Compute Engine instances | Google Cloud Storage pricing | Free egress within regions, | Free tier available for limited | Competitive pricing for compute, strong | Limited enterprise support compared to |

| Cloud Provider | Pricing Model | Compute Costs | Storage Costs | Data Transfer Costs | Free Tier | Key Advantages | Key Disadvantages |
|---|---|---|---|---|---|---|---|
| | Sustained usage discounts | vary by machine type and region | based on usage | charged for inter-region transfer | usage | machine learning integration | AWS |
| **Microsoft Azure** | Pay-as-you-go, Reserved instances, Spot instances | Azure Virtual Machines based on instance size, region | Azure Blob Storage pricing based on storage type and usage | Charges for data transfer out of Azure regions | Free tier available for limited usage | Deep integration with Microsoft software, robust enterprise solutions | Pricing can become high for large-scale deployments |

This table provides a detailed comparison of the cost structures for AWS, Google Cloud, and Microsoft Azure, focusing on the core components of compute, storage, and data transfer costs. It also highlights the key advantages and disadvantages of each cloud provider, helping organizations to make informed decisions based on their specific hybrid cloud deployment needs.

## 5. Compliance and Regulatory Challenges in Hybrid Cloud

### Regulatory Requirements

The move towards hybrid clouds has created need for deep comprehension of compliance and regulatory concerns. Organizations moving their on-premise systems into the public and private cloud infrastructure must make sure that their operations adhere to certain legal, ethical, and compliance standards. As stated by Buyya et al. (2008) and Mell & Grance (2011), the most challenging issue concerning hybrid cloud systems is the compliance regulations that differ for every industry and region. Compliance regulations entail a myriad of issues, including but not limited to, where data is stored and accessed from, privacy issues, and even the levels of openness required in data processing. They include the healthcare industry, the financial sector, and even governmental regulatory bodies that all implement strict rules regarding data compliance like HIPAA in the United States and GDPR in the European Union that specify how data will be handled, processed, and stored within institutions.

As previously highlighted, cloud providers must ensure that their infrastructure complies with defined standards and usually provide certain functionalities to assist organizations in meeting compliance regulations. From the previous discussions, however, it can be inferred that the liability for compliance regulations is not assigned solely to the cloud providers. Organizations that have adopted hybrid cloud environments must also have safeguards and measures in place to ensure that their data handling activities are compliant with legal obligations. This can involve the implementation of encryption, identity controls, along with periodic audits on the compliance status against applicable regulations. Sultan (2009) and Jackson and Jeffery (2015) expand upon the debate surrounding compliance with hybrid clouds. They claim that having multi-cloud environments with disparate security policies and compliance requirements introduce a multi-layered compliance gap that needs to be managed.

Moreover, the jurisdictions from which these cloud solutions operate also constitute a problem. These cloud solutions are predicated on storing and processing data in different regions of the world. Such actions can obscure the boundaries concerning legal jurisdiction in such regions. Organizations must actively track and adapt to the legal and regulatory environment because failure to comply can lead to drastic financial damage, legal complications, and reputational damage.

**Concerns Over Privacy and Data Sovereignty**

The legal issues related to storage and processing of data makes data sovereignty an urgent issue with hybrid cloud adoption. Because data is stored in multiple jurisdictions which are often located in different countries, the organization has to navigate through complex layers of data privacy legislations regarding access, transfer, and protection of the data. Bernstein etal. (2009)

highlight issues pertaining to cross-border data storage and access laws and claim data sovereignty challenges take center stage when an organization's data is fragmented across several public and private clouds located in different regions. In some cases, certain countries have stringent laws regarding data retention within the borders of the nation and access or transfer of such data out of the country is highly regulated.

For instance, the General Data Protection Regulation (GDPR) of the European Union is very strict towards the transfer of sensitive private information to outside regions of the EU. Any hybrid cloud operator has to comply with these regulations regardless of where the data is stored or processed, which can prove to be a challenge for global organisations. Such global businesses have to operate in a puzzle where every piece represents a different set of rules. To comply with these directives, organisations have to implement multiple layers of privacy and security protective measures like encryption, anonymization, or even access restrictions for the data in transit between different countries.

Organizations must also anticipate challenges that may arise from data residency laws and be open to answers that question where an organization's data is allowed to legally reside. And indeed, there could be massive penalties and reputational damage as well as litigation for failure to comply with data sovereignty laws. It is important for all these companies to put in place effective approaches to strategically manage compliance with these requirements. Such approaches include the use of cloud services offered by providers located within the bordered region, data segmentation to store sensitive information in the assigned locations, and enforcement of compliance access control and audit logs.

**Compliance Reports and Security Audits**

Owing to the completed nature of compliance and regulatory concerns in a hybrid cloud environment, every organization striving towards legal compliance is left with no alternative - security audits and compliance reports become a factual necessity. A security audit is defined as an evaluation of an organization's security policies, practices, and infrastructure in order to identify security gaps and compliance issues. These audits are often conducted by either external or internal specialists who examine the cloud infrastructure in relation to the rest of the industry's standards, legal requirements, and security protocols to determine compliance.

The need for such security audits is further exacerbated for hybrid cloud systems because organizations need to evaluate their onpremise systems vis-a-vis the cloud systems. Fox et al. (2009) describes compliance reporting for hybrid cloud deployments as the process where reports are designed with the intent to analyze several components of security, for example, the encryption and access control, and assess whether relevant regulatory levels are encountered. These reports are shared among relevant stakeholders which includes internal personnel, external auditors, and regulatory bodies to provide evidence of compliance and outline plans to address security gaps.

Organizations can minimize the risks of any data breaches, or insufficient data protection by performing a security audit. These breaches, when unchecked, can escalate into other legal issues. The compliance report, on the other hand, captures the essence of the audit, as well as the evidence showing the policies enacted concerning the guidelines. An organization's security policies, its resulting control measures for data protection, and the purported protective measures and gaps identified are accepted within these documents.

## 6. Case Study: Practical Uses of Hybrid Clouds

**A Global Company's Use of Hybrid Cloud Services**

The shift towards adopting a hybrid approach by large organizations is attributed to the combination of public cloud services' flexibility and scalability, and other factors such as: control and security of the private cloud infrastructure. One of the successful implementations of a hybrid cloud is the multi-regional deployment of Amazon Web Services (AWS) and Microsoft Azure. Large multinational corporations have a constant need to maintain a distributed cloud infrastructure to serve the needs of their teams and customers who are spread over a large geographical area. The adoption of AWS and Azure enables the organization to create a hybrid cloud solution that is more tolerant of control issues with sensitive information, while still achieving compliance with the region's laws and regulations.

AWS and Azure are recognized as the two top providers of cloud services technology, and both have created extensive ecosystems of tools and services that form hybrid cloud systems. AWS and Azure have their respective strengths; Powerful computing capabilities, storage solutions, and a wide range of services are offered by AWS, while Azure's forte are enterprise integrations of other Microsoft products within an organization. Organizations can rightfully make use of both cloud environments simultaneously with combined platform support that provides high availability and disaster recovery capabilities across several areas. Google Cloud (n.d.) and Microsoft Azure (n.d.) highlight how combined service integration with these systems enables organization infrastructures to shift resources limits boundaries to maximize efficiency, productivity, dependability, and expenses.

Combining two distinct cloud platforms involves several complexities such as securing data, managing systems, and integrating software applications. Most organizations, according to Armbrust et al. (2009) and Varghese & Buyya (2017), face great difficulties in claiming a true hybrid cloud capability as these systems are usually associated with their own architecture, API and management systems. For instance, without sufficient cloud control systems for migrating between AWS and Azure, clouds remain stranded leading to high operational lag and costs. Moreover, implementing security across multiple clouds is even more challenging, therefore forcing organizations to rethink how security integration is done for the two platforms.

Some of these challenges can be addressed by utilizing hybrid cloud management tools that offer single pane of glass management for both AWS and Azure. With these tools, organizations can track workloads, manage resources efficiently, and enforce standard security policies on both clouds. Furthermore, employing cloud-native technologies like containerization and microservices facilitates a modular approach to application deployment which reduces the integration challenge between AWS and Azure. Table 4 presents a detailed analysis regarding the identified challenges and solutions to this hybrid cloud deployment, highlighting the advantages and disadvantages of merging AWS and Azure for multi-region cloud infrastructure operations (AWS, n.d.; Microsoft Azure, n.d.).

**Lessons Learned and Best Practices**

As noted, deep learning methods that automatically extract perceptively useful visual features from medical images can help develop a model that performs at par with an expert radiologist in identifying pertinent medical evidences that require greater attention. Using such methods on a set of images will yield "fingerprint" features unique to each radiologist marked images which could subsequently be used for the radiologist identification problem. Transfer learning year Quantum neural networks can be integrated with other established deep learning paradigms that include image segmentation and object detection methods to improve accuracy of those methods in the application domains of medical imaging interpreters and.NET programmers.

An organization's best practice in ensuring security across hybrid cloud environments is trust no one and verify every attempt to access resources, regardless of true organizational networks. This approach is highly effective for hybrid cloud implementations where data and applications are hosted with multiple cloud providers and on-premise systems. Moreover, firms must ensure there is uniformity in the policies and controls for the public and private clouds. To illustrate, in the context of a hybrid cloud environment, these risks are reduced through the application of standard encryption, multi-factor authentication, and routine security assessments.

Besides security issues, effective cost and performance management is another prerequisite that has to be achieved for success in hybrid cloud deployment. In a hybrid cloud deployment, both cost and performance optimization is difficult to achieve because organizations have to combine the use of on-premise infrastructure as well as public cloud resources. One effective technique in relation to cost control for hybrid usage is the adoption of a consumption-based pricing approach where the customer is only charged for the actual resources used. This strategy allows organizations to increase or decrease the usage of cloud resources without incurring large capital expenditures on hardware or infrastructure.

Furthermore, there is the possibility of improving performance through cloud optimization tools that manage the scaling of workloads in real-time. Such tools should help mitigate over-provisioning, which is costly, while assisting in ensuring that critical applications receive the requisite resources to function optimally during peak hours.

**7. Proposed Framework For Hybrid Cloud Security and Cost Optimization**

 **A Unified Framework For Hybrid Cloud Management**

To take advantage of public cloud capabilities and private cloud infrastructure, an increasing number of organizations are utilizing hybrid models. Such models pose an issue due to the lack of an overarching approach to security, cost and efficiency optimization. These elements need to be addressed in an integrated manner to derive the full value of hybrid cloud environments. In terms of the integrating security, cost, and performance optimization endeavors, organizations should counterbalance trade-offs between benefits and challenges due to the inherent complexity and fragmentation of hybrid cloud deployment.

Gai et al., 2020, and Reinders, 2020, assert that a unified hybrid cloud management framework provides a solution to all-level automation issues prevalent in the hybrid cloud system. Such frameworks are expected to provide automated policies capture for orchestration, security, and cost monitoring for the integrated systems of the on-premise infrastructure and the public cloud, in addition to the automated management systems. In addition to cloud management policies, organizations need to have robust policies for task allocation that take into account the specific application requirements, tolerance for latency, and sensitivity to security of various workloads within the organization. The modern hybrid cloud paradigm permits scaling of the workload automatically per the realtime requirement while allowing for satisfactory service levels at minimal cost.

Automating the merging of these components is useful to an organization as it simplifies performance and cost management by combining them within one integrated system and enhancing resource productivity while reducing operational costs and the risk of security policy violations. This improvement also helps in the management of the hybrid clouds by ensuring that the resources allocated address the technical prerequisites and business objectives in an optimal manner.

**Suggested policy: Zero Trust, Encryption, and Hybrid Cloud Security Structure.**

Companies that are moving to a hybrid cloud model need to drive change while at the same time managing risk. The boundary for security has fundamentally changed from the epoch of single organization to multi-cloud environment. As the title implies, a zero trust approach does not automatically trust any user, external or internal to the corporate network which makes it perfect for hybrid setups.

 Any request for access has to be verified for each attempt which assists in preventing data from getting into the hands of unauthorized users.

The Strategy of Zero Trust was introduced by Kindervag in 2010 to outline a threat minimization strategy that integrates strong verification by defining boundaries and checkpoints at every access point. Therefore, requesting access to hybrid cloud resources would require rigorous identification, authentication, authorization, and encryption so that exposed sensitive information can be secured. This architecture should be adopted especially for hostile advanced hybrid systems, where the degree of segmentation between public and private clouds determines the level of security against unauthorized access.

Data transfer between an on-premise system and a cloud system requires encryption to keep the data secure. Robust protocols put in place for encryption ensure that there is no unauthorized access or manipulation of data regardless of whether it is being processed, at rest or in transit. Hybrid cloud deployments benefit from additional encryption security because it protects data across multiple endpoints while reducing the chance of the data being breached. Standards for encryption need to be the same for both the private and public clouds as security measures need to be identically applied to the information no matter the location.

Using encryption and implementing a zero trust approach enables an organization to create hybrid cloud policies that regulate the access, usage, and security audit of data. Both public and private clouds are covered under these policies that control which users and devices can access certain resources and outline how to respond to security breaches. Buyya et al. (2008) argue that creating strong hybrid cloud policies is crucial for secure operational capability, especially where there is sensitive or controlled data. Organizations should also ensure that these policies provide direction for such security compliance audits so organizations are able to follow set guidelines and rules of the relevant industry.

The recommended framework highlights major components, for example, identity and access management (IAM), encryption policies, multi-factor authentication (MFA), and active threat monitoring.